\documentclass{article}

    \PassOptionsToPackage{numbers, compress}{natbib}

    \usepackage[final]{neurips_2020}

\usepackage[utf8]{inputenc} 
\usepackage[T1]{fontenc}    
\usepackage{hyperref}       
\usepackage{url}            
\usepackage{booktabs}       
\usepackage{amsfonts}       
\usepackage{nicefrac}       
\usepackage{microtype}      
\usepackage{soul}           
\usepackage{graphicx}
\usepackage{xcolor}
\usepackage{subcaption}
\usepackage[english]{babel}
\title{Learning to estimate a surrogate respiratory signal from cardiac motion by signal-to-signal translation}

\author{%
Akshay Iyer, Clifford Lindsay, P. Hendrik Pretorius, Michael A. King
\\Dept. of Radiology, UMass Medical School, Worcester, MA, USA \\
\{akshay.iyer2, clifford.lindsay, hendrik.pretorius, michael.king\}@umassmed.edu
}

\begin{document}

\maketitle

\begin{abstract}
In this work, we develop a neural network-based method to convert a noisy motion signal generated from segmenting rebinned list-mode cardiac SPECT images, to that of a high-quality surrogate signal, such as those seen from external motion tracking systems (EMTs). This synthetic surrogate will be used as input to our pre-existing motion correction technique developed for EMT surrogate signals. In our method, we test two families of neural networks to translate noisy internal motion to external surrogate: 1) fully connected networks and 2) convolutional neural networks. Our dataset consists of cardiac perfusion SPECT acquisitions for which cardiac motion was estimated (input: center-of-count-mass - COM signals) in conjunction with a respiratory surrogate motion signal acquired using a commercial Vicon Motion Tracking System (GT: EMT signals). We obtained an average R-score of 0.76 between the predicted surrogate and the EMT signal. Our goal is to lay a foundation to guide the optimization of neural networks for respiratory motion correction from SPECT without the need for an EMT.
\end{abstract}

\section{Introduction}
In Single-Photon Emission Computed Tomography (SPECT) and other imaging modalities, uncorrected respiratory motion can lead to image degradation causing blurring of important structures and artifactual distortions. A common method for detecting, measuring, and correcting for respiratory motion-induced artifacts is the usage of External Motion Tracking (EMTs), such as using marker-based motion tracking systems to measure an external surrogate \cite{pret01}. Despite being accurate and reliable for this task, EMTs can incur significant monetary cost, effort, and slow down patient workflows thereby preventing wide adoption in clinical settings.
\par As an alternative to EMTs, our group has recently developed an approach for estimating the motion of the heart directly from cardiac perfusion SPECT acquisitions by segmenting the voxels within a region of interest around the heart, then calculating the center-of-count-mass (COM) in the axial direction, sequentially at short intervals (100ms) over the duration of the entire scan \cite{pret02}. Unfortunately, such short timespans result in noisy cardiac motion signals, which are often noisier than the motion signals produced by EMTs. Furthermore, our group has invested significant effort in developing EMT-based correction and evaluation schemes for cardiac imaging, which we would like to leverage. Therefore, we are laying the foundation for a correction method based on neural networks to perform signal-to-signal translation (i.e., COM to EMT) thereby alleviating the need for expensive EMT hardware and technologist effort but retaining the correction methods developed by \cite{pret01}.

\section{Materials and Methods}
\label{data}

\begin{figure*}[htbp]
  \begin{subfigure}[b]{0.5\linewidth}
    \includegraphics[scale=0.133]{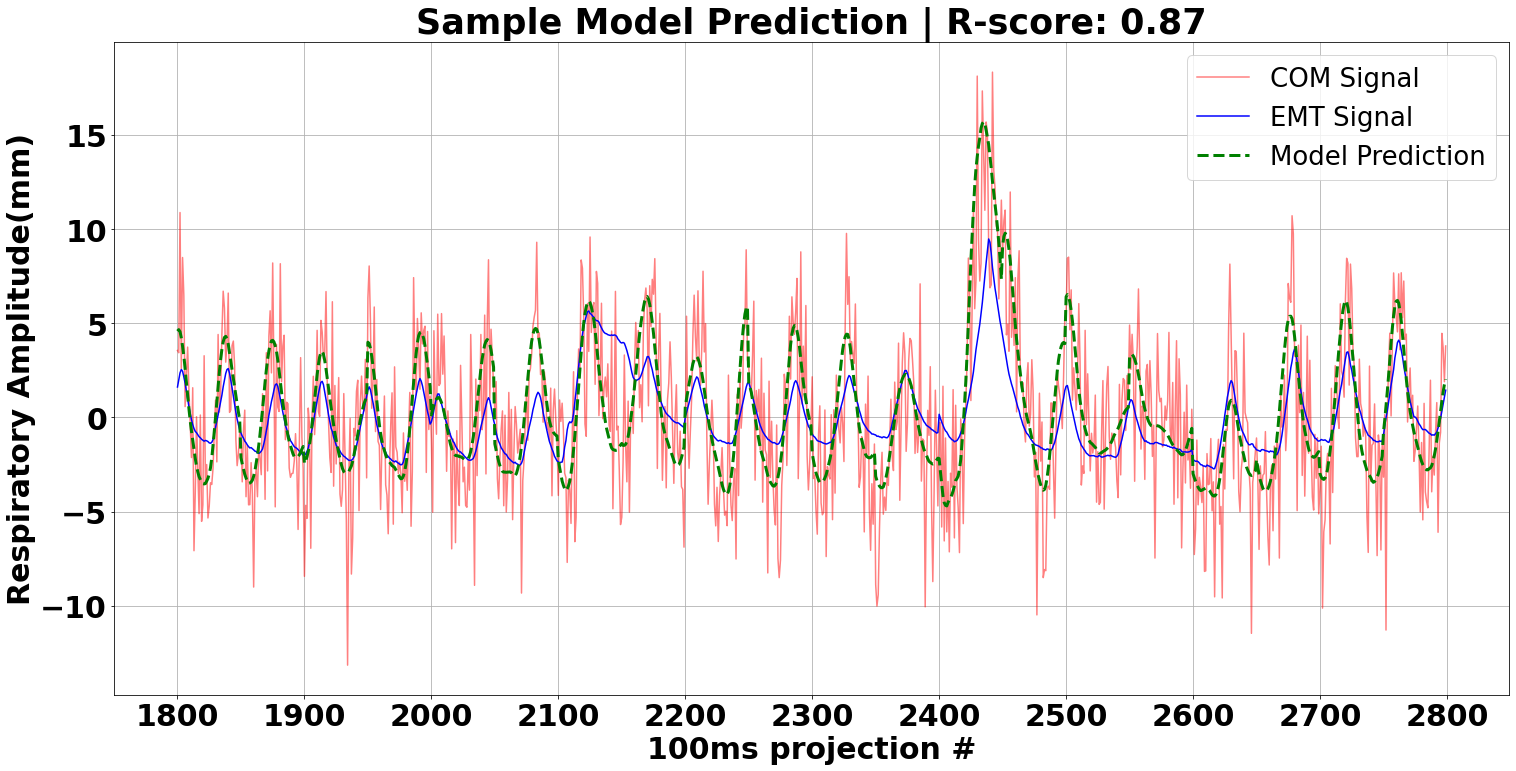}
    \caption{Model prediction with a high R-score}
    \label{goodresult}
  \end{subfigure}
  \begin{subfigure}[b]{0.5\linewidth}
  \includegraphics[scale=0.178]{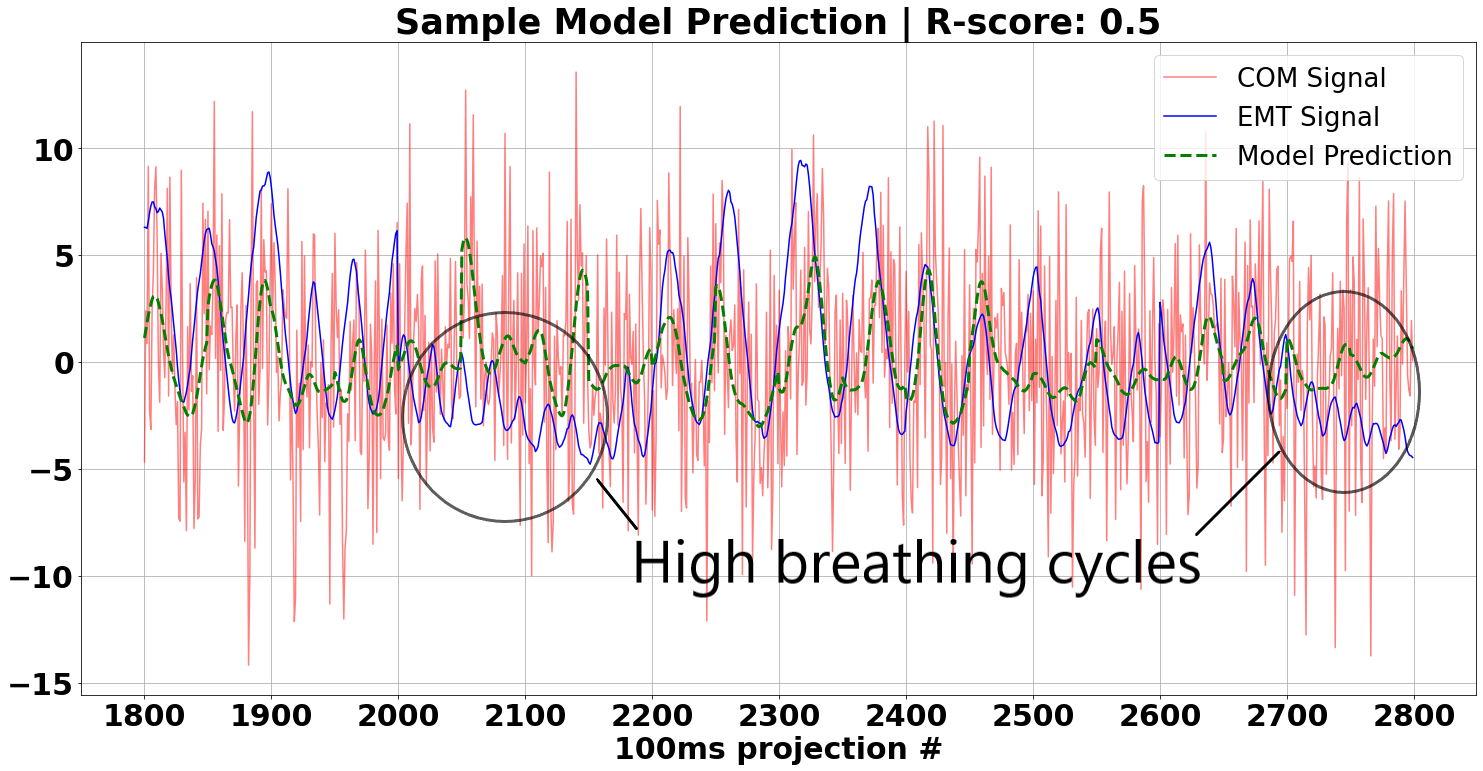}
    \caption{Model prediction with a low R-score}
    \label{badresult}
  \end{subfigure}
  \caption{\textit{Portion of COM (red), EMT (blue), and Surrogate Prediction (green) signals.} Fig. \ref{goodresult} shows the model predicting a 
  surrogate closely matching the EMT signal. Fig.\ref{badresult} shows a model prediction not matching the EMT, especially where there is rapid breathing (circled section) in EMT signals.}
  \label{model_results}
\end{figure*}

\textbf{Center-of-Mass (COM) Signal and External Motion Tracking (EMT) Signal Data: }For this work, we have 200 COM and corresponding EMT signals from an existing dataset. The COM signals were obtained from cardiac perfusion SPECT scans using a 2-headed Philips BrightView XCT SPECT/CT system. Each scan consists of projections from each of its 2 heads at 32 different angles (64 projections in total) rotating around the patient over 180 degrees. Each projection is divided into 200 time intervals of 100ms each (approximately 20s/angle) resulting in a COM signal of a total length of 6400 intervals.  A commercial marker-based Visual Tracking System (Vicon) was used to track the motion of the patient’s abdomen to provide an EMT signal of the same length as the COM signal. Figure \ref{model_results} shows examples of a portion of a sample COM (red line) and EMT (blue line) signals adjusted to have a zero-mean displacement. As can be seen, both signals are highly correlated but the EMT signal has significantly less noise than the COM signal.

\vspace{-3pt}

\textbf{Fully Connected Networks: } Our initial network architecture was a 1-layer NN containing 200 hidden units and was trained to translate COM to a surrogate signal of the same length. The dataset was split into a training and a validation set. For training, both the COM (input) and the EMT (ground truth) signals for each patient was divided into lengths of 200, yielding 8 sub-signals per patient. Mean Squared Error (MSE) loss was used to train the network, but the main metric of interest was a Pearson Correlation Coefficient (R-score) between the predicted surrogate signal and the EMT signal measuring their similarity.
The R-scores were calculated during validation and for the full-length signals. We observed that our fully connected network with 200 hidden units was overfitting. Therefore, a grid-search  method was used to find an optimal number of hidden units and layers. We thus determined that a 1-layer NN with 20 hidden units alleviated any overfitting. Furthermore, a hyper-parameter search over the learning rate, optimizer parameters, input data normalization, and smoothing of input data showed that the best validation R-scores could be obtained using an Adam optimizer with a learning rate of 3e-3. We also determined that neither smoothing nor normalization of input data resulted in a significant improvement in the R-scores. 
Further, it was observed that dividing the signals into smaller sub-signals of length 50 slightly increased the validation R-scores.


\textbf{Convolutional Neural Networks: }
The second network architecture we developed consisted of two different configurations of Convolutional Neural Networks (CNNs) in the form of U-Net \cite{unet} style encoder-decoder networks. As with the fully-connected network, MSE was used as the loss function and the R-score between the COM signal and the EMT signal was used to measure network performance in both cases. Validation R-scores were measured on full-length data for each patient and the optimizer Adam, with a learning rate of 3e-3 showed to perform best in both cases. 
\par For the first network architecture, we converted the 1D signals into 2D representations, to preserve the originally published U-Net network structure and allowing us to perform an image to image translation of the signals. The COM and the EMT signals were divided into sub-signals of length 1024 which were then reshaped as a 2D image of size 32*32. The single-channel 2D image corresponding to the COM data was fed as the input to the U-Net and the single-channel 2D image corresponding to the EMT signal was used as the ground truth. We varied the encoders using \cite{densenet, resnet, xception, dpn, vgg, inception, efficient} and \cite{densenet, resnet} gave the best results. For the second CNN network architecture, a U-Net style encoder-decoder was employed but converted to train on 1D signals of length 1000 as input instead of 2D images as done on the initial CNN architecture. We also varied the network architecture to include Squeeze-and-Excitation (SE) blocks \cite{se_net}, which models interdependencies between channels to improve the representational power of the network. We observed that the U-Net achieved the highest accuracy when it incorporated SE blocks.

\section{Results and Discussion}
\label{results}

\begin{table}
  \caption{Validation R-scores of surrogate signals with the EMT signals for the different models after a 5-fold cross-validation. We observed the best performance with U-Net and Squeeze-and-Excitation blocks.}
  \label{scores}
  \centering
  \scriptsize
  \begin{tabular}{ccccc}
    \toprule
    FullyConn-200 & FullyConn-50 & U-Net(2D) & ModelAvg & U-Net (1D+SE)  \\
    \midrule
     $0.67$ & $0.72$ & $0.70$ & $0.73$ & $0.76$    \\
    \bottomrule
  \end{tabular}
\end{table}

\begin{figure*}
  \begin{subfigure}[b]{0.5\linewidth}
    \includegraphics[height=0.15\textheight, keepaspectratio]{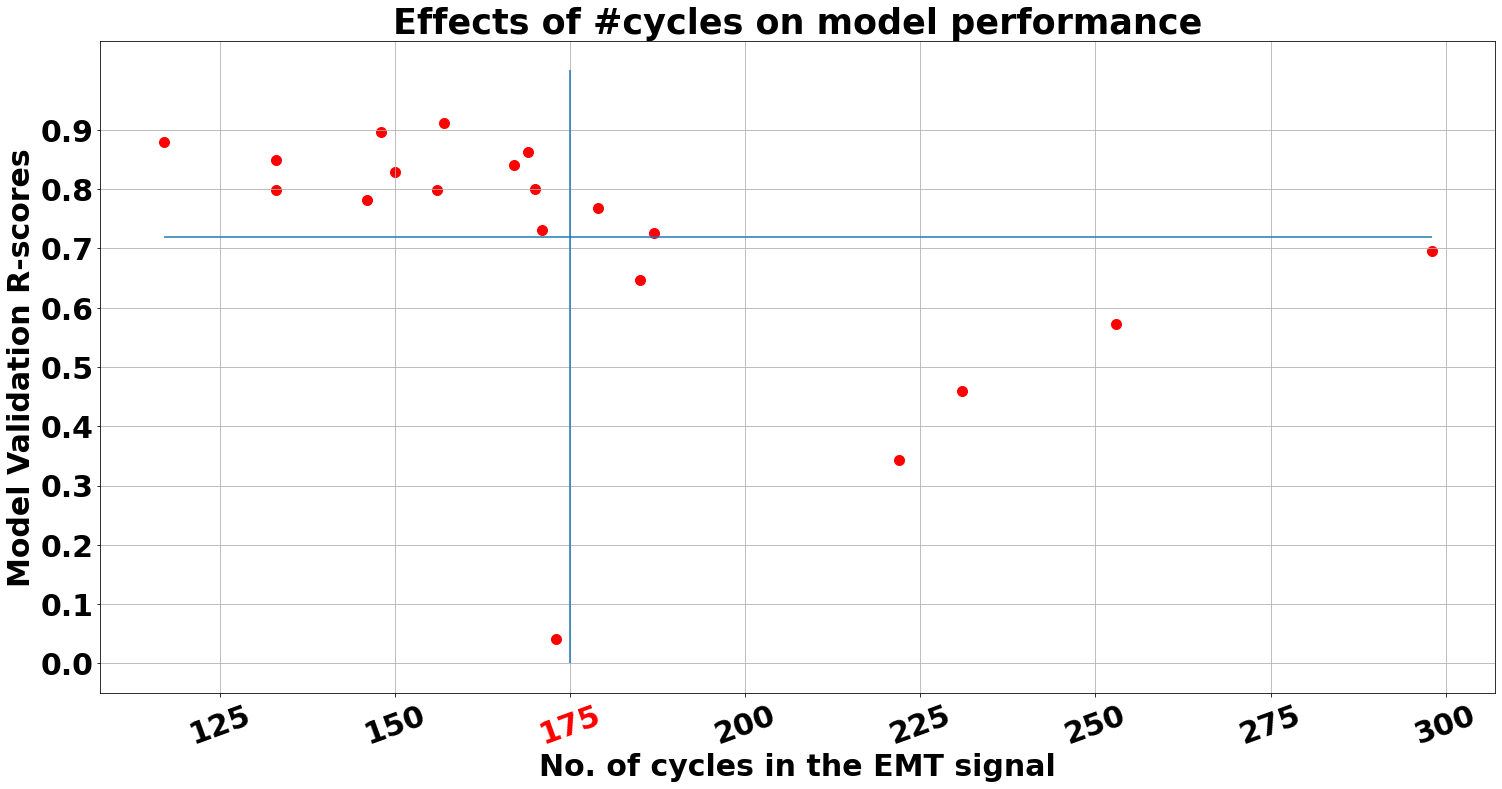}
    \caption{Effect of breathing cycles on model performance}
    \label{cycles}
  \end{subfigure}
  \begin{subfigure}[b]{0.5\linewidth}
    \includegraphics[height=0.13\textheight, keepaspectratio]{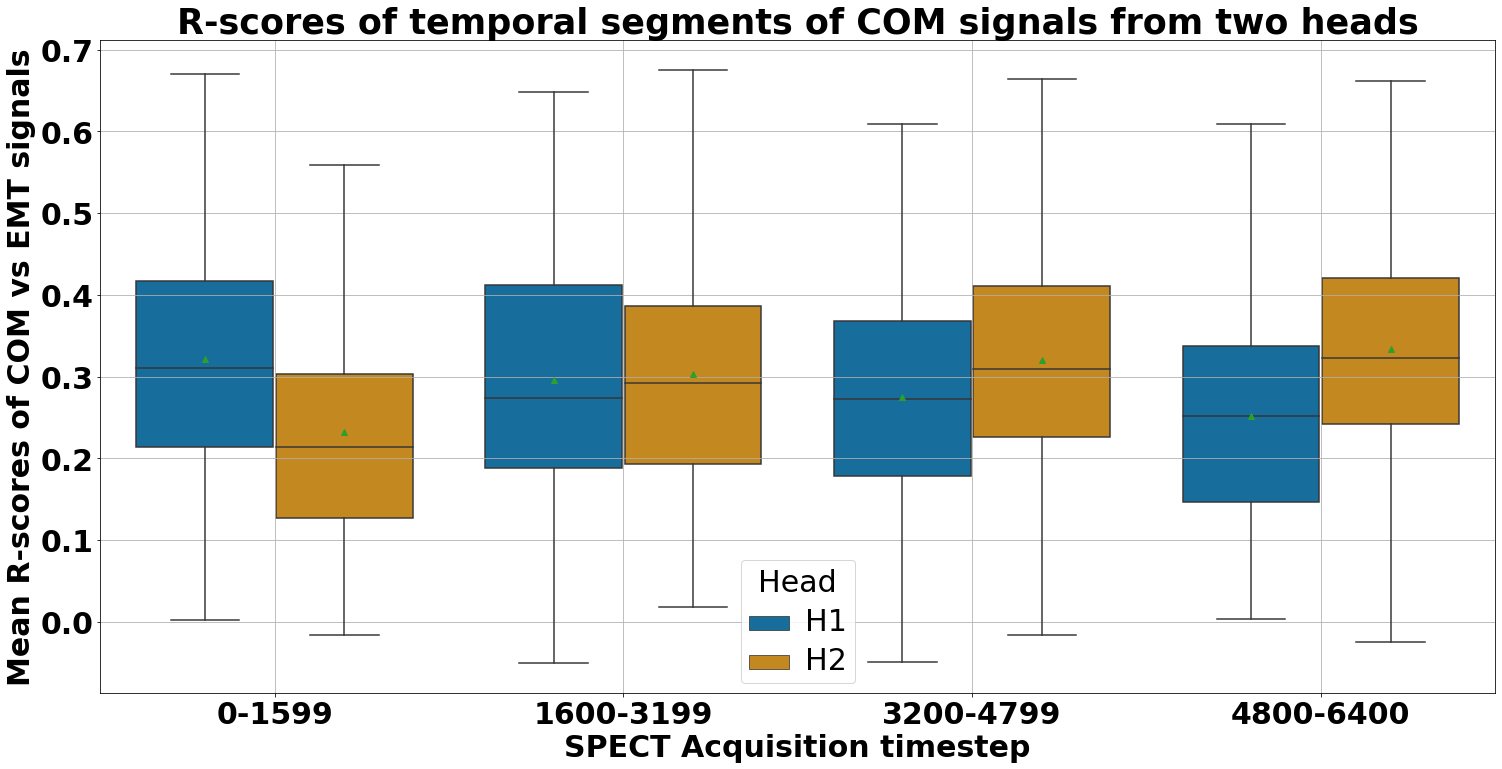}
    \caption{Difference in R-scores of signals from 2 camera heads}
    \label{2heads}
  \end{subfigure}
  \caption{Fig. \ref{cycles} shows that the model performance is above average when there are less than 175 breathing cycles in a 6400-length signal. Fig. \ref{2heads} shows that the signal of the 2nd head produces a lower score than the 1st head, during the 1st quarter while the opposite is true in the last quarter.}
  \label{analysis}
\end{figure*}

Table \ref{scores} shows the validation R-scores obtained after a 5-fold cross-validation. We observed that using the U-Net with a 1-D signal input and Squeeze-and-Excitation blocks obtained the highest R-scores. Next, we observed that the fully connected network performs better with smaller signals of length (50) than longer signals (200). U-Net with a 2D input works well when included in an ensemble with the fully-connected network (FullyConn-50). Furthermore, on inspection of the network predictions, it was observed that some of the predicted surrogate signals had R-scores of 0.85 and above, while a different group’s predictions had lower R-scores which was bringing down the average (see Fig. \ref{model_results}). An investigation revealed model performance was tied to the number of breathing cycles in the EMT signal and the original R-score of the COM signals with the EMT signals. One can clearly see in Fig. \ref{cycles} that, on average, if the COM signal has more than 175 breathing cycles, then the model performance is below average. This effect can also be observed in Fig. \ref{model_results}, where the EMT signal has a higher number of breathing cycles in Fig \ref{badresult} than in Fig. \ref{goodresult}, resulting in a lower R-score.

\vspace{-3pt}

\par We also observed that all the model predictions tended to be above average if the R-scores of the original COM signal and the EMT signal were above 0.35. This implies that our model prediction, regardless of model type, was highly dependent on the original R-score. We further observed that the second and third quarters of the COM signal (time steps 1600-3200 and 3200-4800) have a higher R-score with the EMT signal across all patients While the second and third quarters of the signal have a mean R-score with EMT signal across the patients of 0.39, the first quarter has an R-score of 0.36, and the last quarter has around 0.38. Since the COM signal is a combination of signals from two different camera heads, we analyze the R-scores of the signals from each head separately with the EMT signal as shown in Fig. \ref{2heads}. Here, we observe that in the first and the last quarter, there is a difference of over 0.1 between the R-scores of the two heads thus lowering the R-score of the combination. This temporal difference between the R-scores is on account of the position of the two SPECT camera heads during different time steps of the scan. 
Using this knowledge, we created a separate dataset of COM and EMT signals which contained samples deemed more difficult based on the R-scores and the number of breathing cycles present. This separate dataset was used to mitigate their effects using strategies like training on the entire dataset including the difficult samples, and then trained on just the difficult samples, or interleaving batches of regular and difficult data samples, or curriculum learning \cite{curr} by presenting samples in increasing order of difficulty. While these strategies resulted in minor improvements in performance, they present potential avenues for further investigation and we’ll continue to investigate them.
\par Thus, in this work, we lay a foundation for deep learning-based respiratory motion estimation in SPECT and identify key areas, which could be focused upon, to further improve estimation accuracy.

\begin{ack}
Research reported in this publication was supported by the National Heart, Lung, and Blood Institute of the National Institutes of Health under Award Number R01 HL154687. The content is solely the responsibility of the authors and does not necessarily represent the official views of the National Institutes of Health.
\end{ack}







\setcitestyle{numbers}
\bibliography{refs}

\end{document}